\documentclass{article}
\usepackage{article_defaults}

\newcommand{\jvrev}[1]{#1}

\toptitle{A Paraboloid Fitting Technique for Calculating Curvature from Piecewise-Linear Interface Reconstructions on 3D Unstructured Meshes}

\author[1]{Z. Jibben\thanks{Corresponding author: zjibben@lanl.gov}}
\author[1]{N. N. Carlson}
\author[2]{M. M. Francois}

\affil[1]{Los Alamos National Laboratory, CCS-2 Computational Physics and Methods, Los Alamos, NM 87545, USA}
\affil[2]{Los Alamos National Laboratory, XCP-4 Methods and Algorithms, Los Alamos, NM 87545, USA}

\date{\today}
\abstract{
  We present a novel method for calculating interface curvature on 3D unstructured meshes from piecewise-linear interface reconstructions typically generated in the volume of fluid method. \jvrev{Interface curvature is a necessary quantity to calculate in order to model surface tension driven flow. Curvature needs only to be computed in cells containing an interface.} The approach requires a stencil containing only neighbors sharing a node with a target cell, and calculates curvature from a least-squares paraboloid fit to the interface reconstructions. \jvrev{This involves solving} a $6\times6$ symmetric linear system in each mixed cell. We present verification tests where we calculate the curvature of a sphere, an ellipsoid, \jvrev{and a sinusoid} in a 3D domain \jvrev{on regular Cartesian meshes, distorted hex meshes, and tetrahedral meshes. For both regular and unstructured meshes, we find in all cases the paraboloid fitting method for curvature to converge} between first and second order with grid refinement.
}

\begin{document}
\maketoptitle

\section{Introduction}
The volume of fluid (VOF) method \autocite{rider_reconstructing_1998, hirt_volume_1981} is a popular approach to track immiscible material interfaces in fluid simulations, and is noted for its ability to guarantee mass conservation. However, compared to level set or front tracking methods it is notorious for the greater difficulty in calculating geometric properties, such as interfacial curvature \jvrev{required to model surface tension}, particularly on unstructured meshes. There are a variety of methods for calculating curvature from volume fractions, as surveyed by \textcite{cummins_estimating_2005}. However, these approaches are hindered either by \jvrev{relying on a regular mesh structure, requiring a large stencil, or failing to converge under grid refinement.}

For instance, the height function method produces second order curvature calculations, but requires a large computational stencil, \jvrev{typically extending three cells away from a target cell in order to calculate mean interface heights,} increasing communication \jvrev{complexity and} overhead in parallel simulations. It also typically requires regularly spaced data. It has been adapted to AMR meshes \autocite{popinet_accurate_2009} and unstructured meshes \autocite{ivey_accurate_2015}, in both cases by projecting the volume fraction data into a local regular mesh surrounding each target cell. \textcite{francois_interface_2010} developed an approach to use it for non-uniform 2D rectangular meshes. \jvrev{The height function can produce fourth order accurate curvature by increasing the stencil to include five columns rather than three for second order \autocite{figueiredo_high-order_2007,francois_interface_2010}.} \textcite{ito_high-precision_2014} introduced a method to use it directly on 2D unstructured meshes, though they were limited to about first order accuracy.  \textcite{owkes_mesh-decoupled_2015} developed a mesh-decoupled height function method in the context of regular Cartesian grids, which may be extended to unstructured grids. Two other popular methods for calculating curvature, the reconstructed distance function method and the convolved VOF, are compatible with unstructured meshes, though both also require large stencil sizes and have been found to lack convergence with grid refinement \autocite{cummins_estimating_2005,ito_high-precision_2014,ivey_accurate_2015}.

A different approach to curvature calculation is to reconstruct a higher order interface, from which curvature is readily available.
The PROST method \autocite{renardy_prost:_2002} is one of these, which performs a least-squares fit of an implicit quadratic interface to volume fraction data in Cartesian cells. \textcite{evrard_estimation_2017} extended this concept to 2D unstructured meshes. The benefit to these approaches is that the higher order interface reconstruction, and curvature, are calculated with a small stencil consisting only of nearest neighbors. However, they both require iterative minimization algorithms, where at each iteration computational cells must be intersected with a quadratic surface.
Another approach involves fitting a paraboloid surface to a neighborhood of points found from the interface reconstruction centers of mass \autocite{aulisa_interface_2007, scardovelli_interface_2003,bogner_curvature_2016}.
\textcite{denner_fully-coupled_2014} presented a method to calculate a least-squares quadratic fit to volume fraction data, from which curvature is calculated.
\textcite{popinet_accurate_2009} utilized a technique to fit a paraboloid to height function data for curvature calculation as a fallback option when direct use of the height function method fails.

\jvrev{We propose an} approach to fit a surface in a volume-matching sense \jvrev{to the piecewise-linear interface calculations (PLIC) already generated as part of a VOF-PLIC approach,} suitable for direct use on a 3D unstructured mesh. To accomplish this, we attempt to find an analytic fit which approximates a collection of neighboring interface reconstructions in a least-squares sense. This leads to a \jvrev{small} linear minimization problem \jvrev{on each mixed cell}, which is solved \jvrev{directly}. From this analytic fit, curvature calculation is straightforward.
In \secref{geometry} we describe the basic geometric algorithms we utilize in this paper. \secref{curvature} describes our algorithm for generating paraboloid surface fits from interface reconstructions, as well as the curvature calculation from that surface. Finally, in \secref{results} we show verification tests and results for this algorithm.

\section{Numerical Methods}
\jvrev{We are interested in calculating interface curvature within
  a VOF-PLIC framework on an unstructured mesh. At a given time step, we are given a mesh of arbitrary polygon cells, and with each cell is associated a volume fraction $\alpha$.
  The volume fraction field indicates in each cell what fraction of volume is filled by a chosen phase. In each mixed cell (those with $0 < \alpha < 1$) the interface reconstructed as a plane in order to calculate volume fluxes.
  Calculating the interfacial curvature $\kappa$
  from these PLICs is the primary interest of this paper.}

\subsection{\jvrev{PLIC Geometry}} \label{sec:geometry}

\subsubsection{Non-Convex Polyhedron Subdivision} \label{sec:tesselate}
\begin{figure}[tp]
  \centering
  \includegraphics[width=0.7\textwidth]{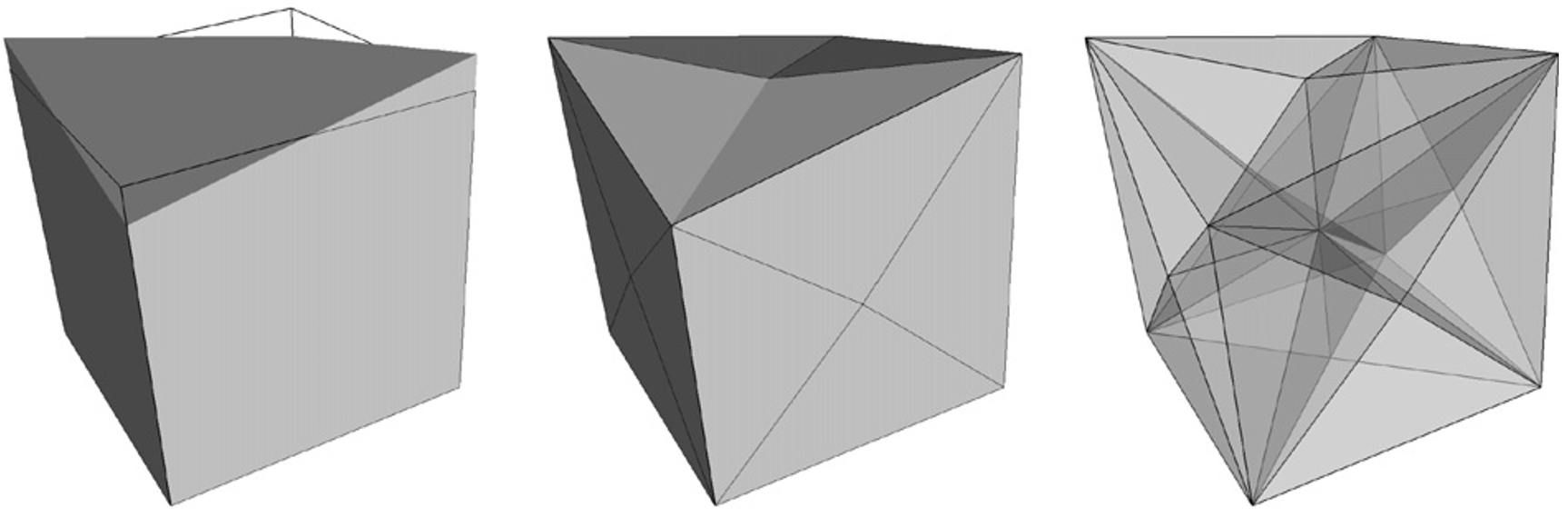}
  \caption{\jvrev{Polyhedron subdivision. Polyhedron with non-planar face shown to the left. Face triangulation is shown in the middle, and sub-division on the right. Figure from \textcite{ahn_multi-material_2007}.}} \label{fig:subdivide}
\end{figure}

\jvrev{For a general unstructured mesh, cells are not guaranteed to be convex, nor are \jvrev{they} guaranteed \jvrev{to have} planar faces. This can present problems particularly for geometric calculations as needed in volume of fluid methods. For instance, a PLIC may provide a plane which intercepts multiple sections of a non-planar cell face. This can occur in any non-regular mesh with cells that have more than three nodes per face. To avoid this problem, we} sub-divide non-convex mesh elements into a collection of tetrahedra, which are defined by introducing new nodes at cell-centers and face-centers,
\jvrev{defined by algebraic averages of node coordinates, as illustrated in \figref{subdivide}.}
This is identical to the approach used by \textcite{ahn_multi-material_2007}.

\subsubsection{Polyhedron-Plane Intersection Polygon}
\jvrev{On mixed cells the interface is reconstructed as a plane which cuts off the appropriate volume in each cell. To achieve convergent curvature results using the approach described in the following section, we find it is necessary to have second order accurate interface reconstructions. Most importantly, this means we require second order normal vectors. Two options with small stencils are the moment of fluid method and LVIRA method, both described for 3D unstructured meshes by \textcite{ahn_multi-material_2007}. For this work, we use the LVIRA method \autocite{pilliod_jr._second-order_2004,ahn_multi-material_2007}.}

\begin{figure}[tp]
  \centering
  \includegraphics[width=0.3\textwidth]{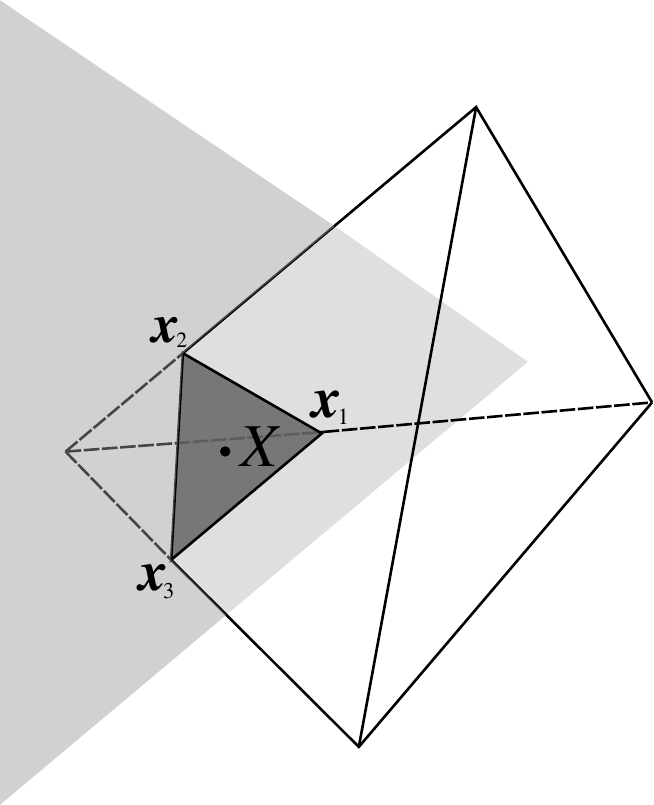}
  \caption{Polyhedron-plane intersection. \jvrev{The planar interface reconstruction is shown in light gray, and the reconstruction polygon in dark gray. The polygon vertices are indicated as $\v{x}_1, \v{x}_2, \v{x}_3$, and the centroid by $\v{X}$.}}
  \label{fig:reconstruction_polygon}
\end{figure}

\jvrev{For our surface fitting technique, we require the polygon provided by the intersection between the planar interface reconstruction and the polyhedral cell,} as depicted in \figref{reconstruction_polygon}. This \jvrev{requires that we find the intersection points between polyhedron edges and the plane, which we then sort such that they are numbered counter-clockwise with respect to the} plane normal vector. The polygon $r$ is then adequately described by a set of points $\{\v{x}_{r,v}\}$. The pair $(\v{x}_{r,v},\v{x}_{r,v+1})$, with $v+1$ appropriately looping back to the first vertex for the final edge, gives an edge of the polygon.

Given a non-convex polyhedron, the cell is split into a collection of tetrahedra as described in the previous section. Then, the polyhedron-plane intersection is given by a collection of polygons, where this calculation is repeated for each sub-tetrahedron.

\subsubsection{Polygon Centroid}
We calculate \jvrev{planar} polygon centroids by
\begin{align}
  \v{X}_r &= \frac{1}{6A_r} \sum_{v=1}^{N_r} \left((\v{x}_{r,v} - \v{c}_r) + (\v{x}_{r,v+1} - \v{c}_r)\right) \left((\v{x}_{r,v} - \v{c}_r) \times (\v{x}_{r,v+1} - \v{c}_r)\right) \cdot \nv{n}_r + \v{c}_r
  \\
  A_r &= \frac{1}{2} \sum_{v=1}^{N_r} \left(\v{x}_{r,v} \times \v{x}_{r,v+1}\right) \cdot \nv{n}_r
\end{align}
where $N_r$ is the number of vertices for this polygon, $A_r$ is the polygon area, and $\nv{n}_r$ is the plane normal vector. \jvrev{The point $\v{c}_r$ may be chosen anywhere in the plane of the polygon; we choose the algebraic average of the polygon vertices.}

\subsection{Calculating Curvature from Interface Reconstructions} \label{sec:curvature}
Essential to any \jvrev{VOF-PLIC} method is reconstructing an interface from volume fraction data. After this has occurred, we have a collection of planes from which we can estimate curvature.

\begin{figure}[t]
  \centering
  \includegraphics[width=\textwidth]{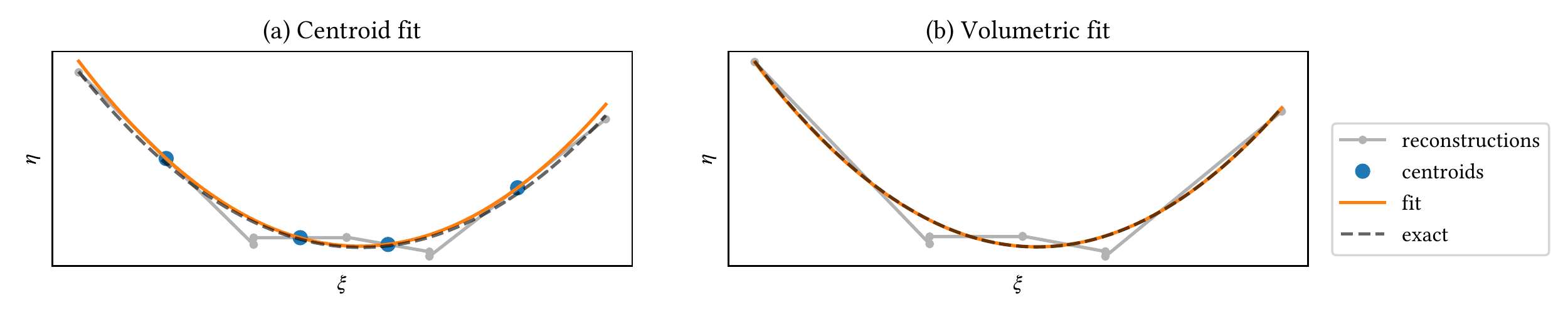}
  \caption{Fitting approaches}
  \label{fig:fits}
\end{figure}

One approach might be to produce a set of points from each interface reconstruction. For instance, we might pick reconstruction polygon centroids (as shown in \figref{fits}a). Then, a surface is fit to these points using a least-squares approach. There exist a variety of methods for fitting a quadric surface to points \autocite{andrews_type-constrained_2014, rangarajan_hyper_2010, taubin_estimation_1991}, as well as direct paraboloid fitting to PLIC \jvrev{centroids} as used by \textcite{scardovelli_interface_2003,bogner_curvature_2016}. However, we find curvature calculations from such approaches not to converge under grid refinement, in line with the findings of \autocite{scardovelli_interface_2003}.

\begin{figure}[t]
  \centering
  \includegraphics[width=0.5\textwidth]{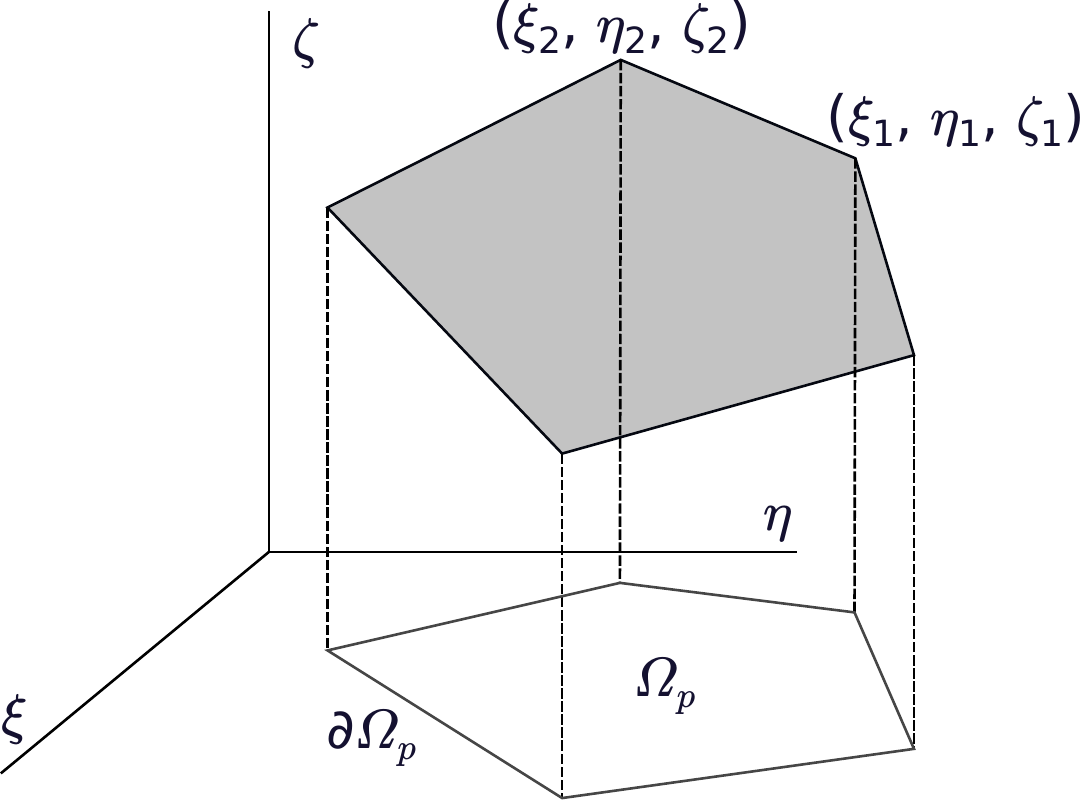}
  \caption{Interface reconstruction polygon integration domain}
  \label{fig:polygon_integral}
\end{figure}

A better approach is \jvrev{suggested} by the fact that interface reconstructions only represent the \jvrev{volume cut off by the interface, so} identifying a set of points which will produce a good surface reconstruction is difficult.
\jvrev{Volume-matching is the basis of several other high order} interface reconstruction techniques \autocite{renardy_prost:_2002,evrard_estimation_2017}, where surfaces are calculated by iteratively matching volume fractions in a collection of cells. Here, we make the assumption that we can use the volumes given by interface reconstructions to avoid intersecting our polyhedral cells with curved surfaces, which leads to a small linear system to be solved in each cell. This concept is illustrated in \figref{fits}b, where we \jvrev{seek} a surface which, for each reconstruction polygon, the volume beneath the interface reconstruction is equal to the volume beneath the curved surface. This concept is done for generally oriented interface reconstruction by first rotating and translating to a coordinate system $(\xi,\eta,\zeta)$, where the $\zeta$ coordinate is aligned with the target cell's interface normal vector $\nv{n}$ and the target cell's reconstruction polygon centroid is \jvrev{located} at $\xi=\eta=\zeta=0$ for convenience. \jvrev{\figref{polygon_integral} shows a general interface reconstruction polygon in this space}.

We define our surface as
\begin{equation}
  \zeta = f(\xi,\eta) = c_0 + c_1 \xi + c_2 \eta + c_3 \xi^2 + c_4 \xi\eta + c_5 \eta^2,
\end{equation}
\jvrev{and} choose the following error function \jvrev{of the paraboloid coefficients $\v{c} = (c_0, c_1, c_2, c_3, c_4, c_5)^{\mathrm{T}}$}, which implies perfect volume matching when it evaluates to zero.
\begin{equation} \label{eq:errorfunc}
  E(\v{c}) = \sum_r \left(\int_{\Omega_r}\! \left(f(\xi,\eta) - \zeta_r\right) \;\mathrm{d}A\right)^2
\end{equation}
Here, the domain $\Omega_r$ refers to the domain of the reconstruction polygon $r$ in the $\xi$-$\eta$ plane, shown in \figref{polygon_integral}. We sum over all $r$ in cells sharing a node with our target cell \jvrev{$r=1$}, in which we wish to calculate curvature. The interface reconstruction here is given by
\begin{equation}
  \zeta_r = b_{r0} + b_{r1} \xi + b_{r2} \eta.
\end{equation}
\jvrev{Note this error function breaks down in the case of under-resolved interfaces, where interface reconstructions neighboring our target cell have $\nv{n}_r\cdot\nv{n}_1 \le 0$. Minimizing \eqref{errorfunc} with respect to the coefficients $\v{c}$ leads to the equations}
\begin{equation} \label{eq:min_sys}
  \sum_r \left(\int_{\Omega_r}\! \left(f(\xi,\eta) - \zeta_r\right) \;\mathrm{d}A\right) \left(\int_{\Omega_r}\! \phi \;\mathrm{d}A\right) = 0
\end{equation}
\jvrev{to be solved for each $\phi\in\{1,\xi,\eta,\xi^2,\xi\eta,\eta^2\}$.}

\jvrev{Eqs.~(\ref{eq:min_sys}) involve} area integrals of the monomial terms over the domain of each interface reconstruction polygon in our patch (depicted in \figref{polygon_integral}). Using Green's theorem, these are transformed into line integrals over the polygon edges, given vertices $(\xi_{r,v}, \eta_{r,v})$ on each polygon. The resulting discrete forms are
\begin{equation} \label{eq:integrals}
  \begin{aligned}
    s_{r0} &= \int_{\Omega_r}     \;\mathrm{d}A = \frac{1}{2} \sum_{v=1}^{N_r} (\xi_{r,v} \eta_{r,v+1} - \xi_{r,v+1} \eta_{r,v}) \\
    s_{r1} &= \int_{\Omega_r} \xi   \;\mathrm{d}A = \frac{1}{6} \sum_{v=1}^{N_r} (\xi_{r,v} + \xi_{r,v+1}) (\xi_{r,v} \eta_{r,v+1} - \xi_{r,v+1} \eta_{r,v}) \\
    s_{r2} &= \int_{\Omega_r} \eta   \;\mathrm{d}A = \frac{1}{6} \sum_{v=1}^{N_r} (\eta_{r,v} + \eta_{r,v+1}) (\xi_{r,v} \eta_{r,v+1} - \xi_{r,v+1} \eta_{r,v}) \\
    s_{r3} &= \int_{\Omega_r} \xi^2 \;\mathrm{d}A = \frac{1}{12} \sum_{v=1}^{N_r} (\xi_{r,v} + \xi_{r,v+1}) (\xi_{r,v}^2 + \xi_{r,v+1}^2) (\eta_{r,v+1} - \eta_{r,v}) \\
    s_{r4} &= \int_{\Omega_r} \xi\eta  \;\mathrm{d}A = \frac{1}{24} \sum_{v=1}^{N_r} (\xi_{r,v}\eta_{r,v+1} - \xi_{r,v+1}\eta_{r,v}) (2\xi_{r,v}\eta_{r,v} + \xi_{r,v}\eta_{r,v+1} + \xi_{r,v+1}\eta_{r,v} + 2\xi_{r,v+1}\eta_{r,v+1}) \\
    s_{r5} &= \int_{\Omega_r} \eta^2 \;\mathrm{d}A = \frac{1}{12} \sum_{v=1}^{N_r} (\xi_{r,v} - \xi_{r,v+1}) (\eta_{r,v} + \eta_{r,v+1}) (\eta_{r,v}^2 + \eta_{r,v+1}^2)
  \end{aligned}
\end{equation}
This leads to a $6\times6$ symmetric linear system for the paraboloid coefficients $\v{c}$,
\begin{equation} \label{eq:linsys}
  \begin{gathered}
    A\v{c} = \v{d} \\
    A_{ij} = \sum_r s_{ri} s_{rj},
    \quad
    d_i = \sum_r s_{ri} (b_{r0} s_{r0} + b_{r1} s_{r1} + b_{r2} s_{r2}).
  \end{gathered}
\end{equation}

This system must be modified for the case where non-convex mesh elements are present. Now, each mixed cell contains a collection of coplanar polygons which represent an interface reconstruction. The paraboloid fit must match the volume given by this polygon set as a whole, rather than match each polygon individually. To accomplish this, our error function is modified to the following
\begin{equation}
  E(\v{c}) = \sum_r \left(\sum_p \int_{\Omega_{rp}} \zeta_r - f(\xi,\eta) \;\mathrm{d}A\right)^2
\end{equation}
Note that this reduces to \eqref{errorfunc} when there exists at most one interface reconstruction polygon per cell. The expressions given by \eqref{integrals} remain unchanged except for referring to the domain $\Omega_{rp}$ and the polygon identified by $r$ and $p$. Since the polygons within a cell are coplanar, the linear function $\zeta_r$ and the coefficients $\v{b}_r$ are independent of $p$. This changes the linear system given by \eqref{linsys} to
\begin{equation}
  \begin{aligned}
    A_{ij} &= \sum_r \left(\sum_p s_{rpi}\right) \left(\sum_p s_{rpj}\right), \\
    d_i &= \sum_r \left(\sum_p s_{rpi}\right) \left(\sum_p (b_{r0} s_{rp0} + b_{r1} s_{rp1} + b_{r2} s_{rp2})\right).
  \end{aligned}
\end{equation}

With the paraboloid coefficients $\v{c}$ given by the solution to this system, it is straightforward to calculate the interface curvature via
\begin{align} \label{eq:parabcurv}
  \kappa &= - \frac{f_{\xi\xi} + f_{\eta\eta} + f_{\xi\xi}f_\eta^2 + f_{\eta\eta}f_\xi^2 - 2f_{\xi\eta}f_\xi f_\eta}{\left( 1 + f_\xi^2 + f_\eta^2\right)^{3/2}}.
\end{align}
We choose to evaluate the curvature at the interface reconstruction centroid, $\xi=\eta=0$.

\section{Results} \label{sec:results}

\begin{figure}[t]
  \centering
  \begin{subfigure}{0.25\textwidth}
    \includegraphics[width=0.95\textwidth]{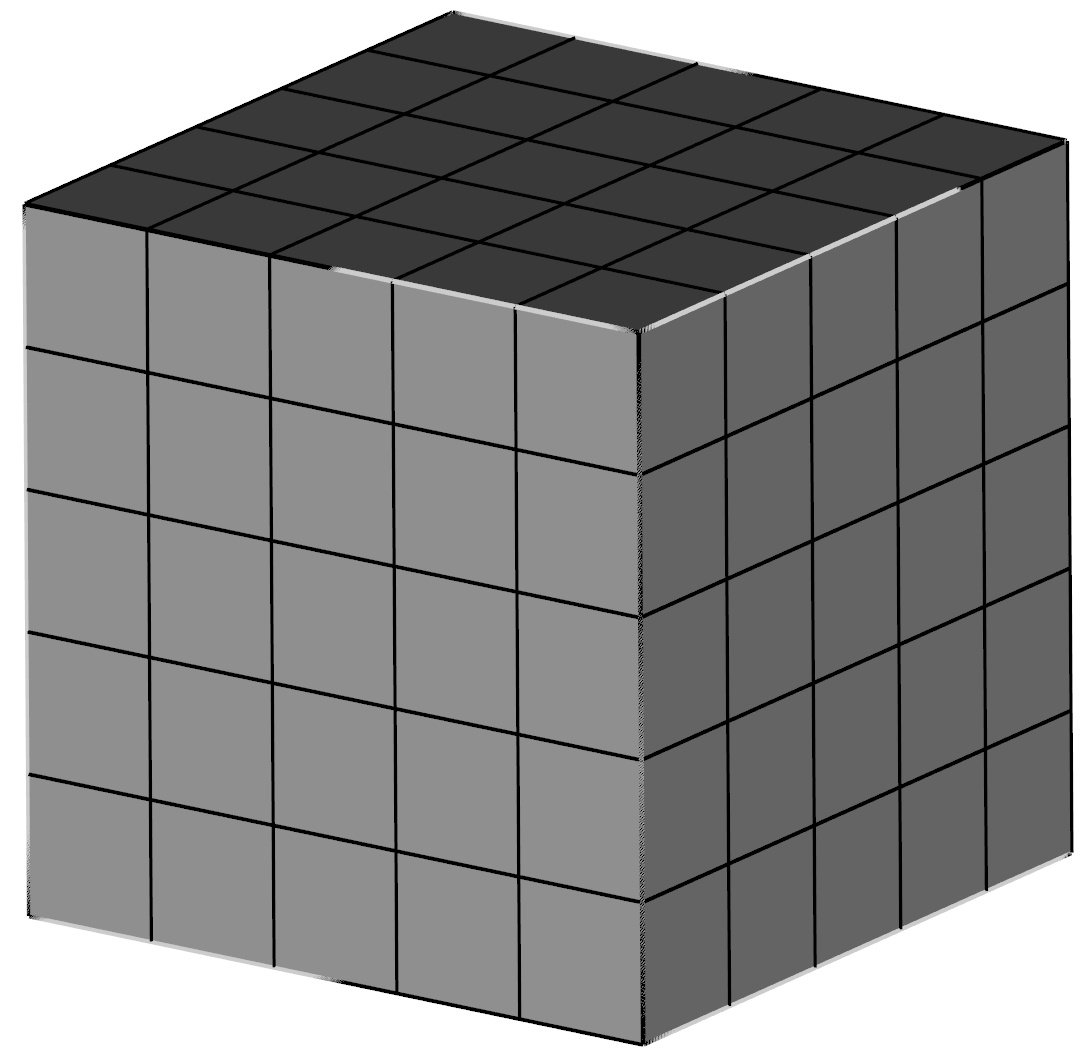}
    \caption{Regular hexahedra}
  \end{subfigure}
  \hspace{1em}
  \begin{subfigure}{0.25\textwidth}
    \includegraphics[width=0.95\textwidth]{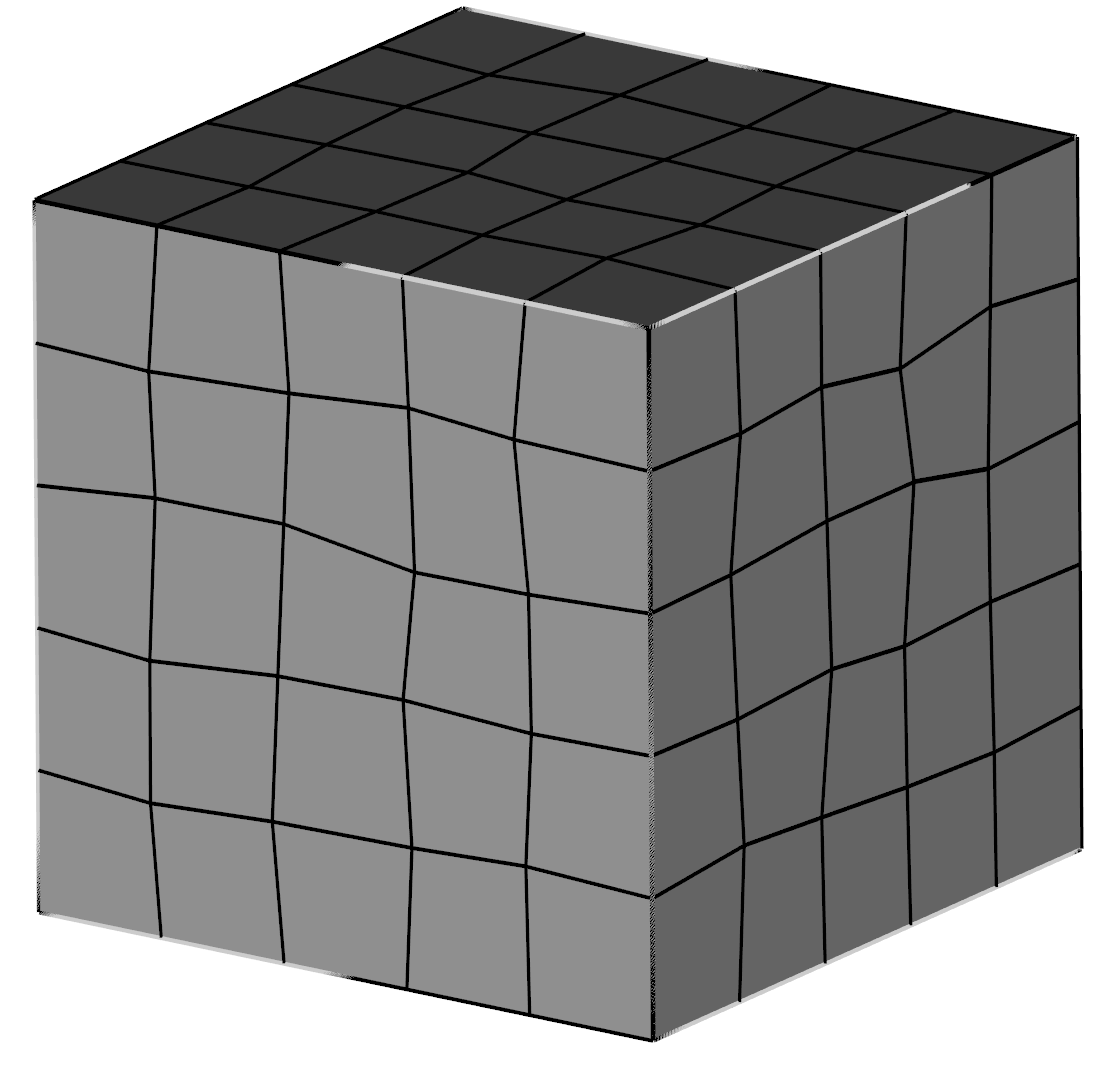}
    \caption{Distorted hexahedra}
  \end{subfigure}
  \hspace{1em}
  \begin{subfigure}{0.25\textwidth}
    \includegraphics[width=0.95\textwidth]{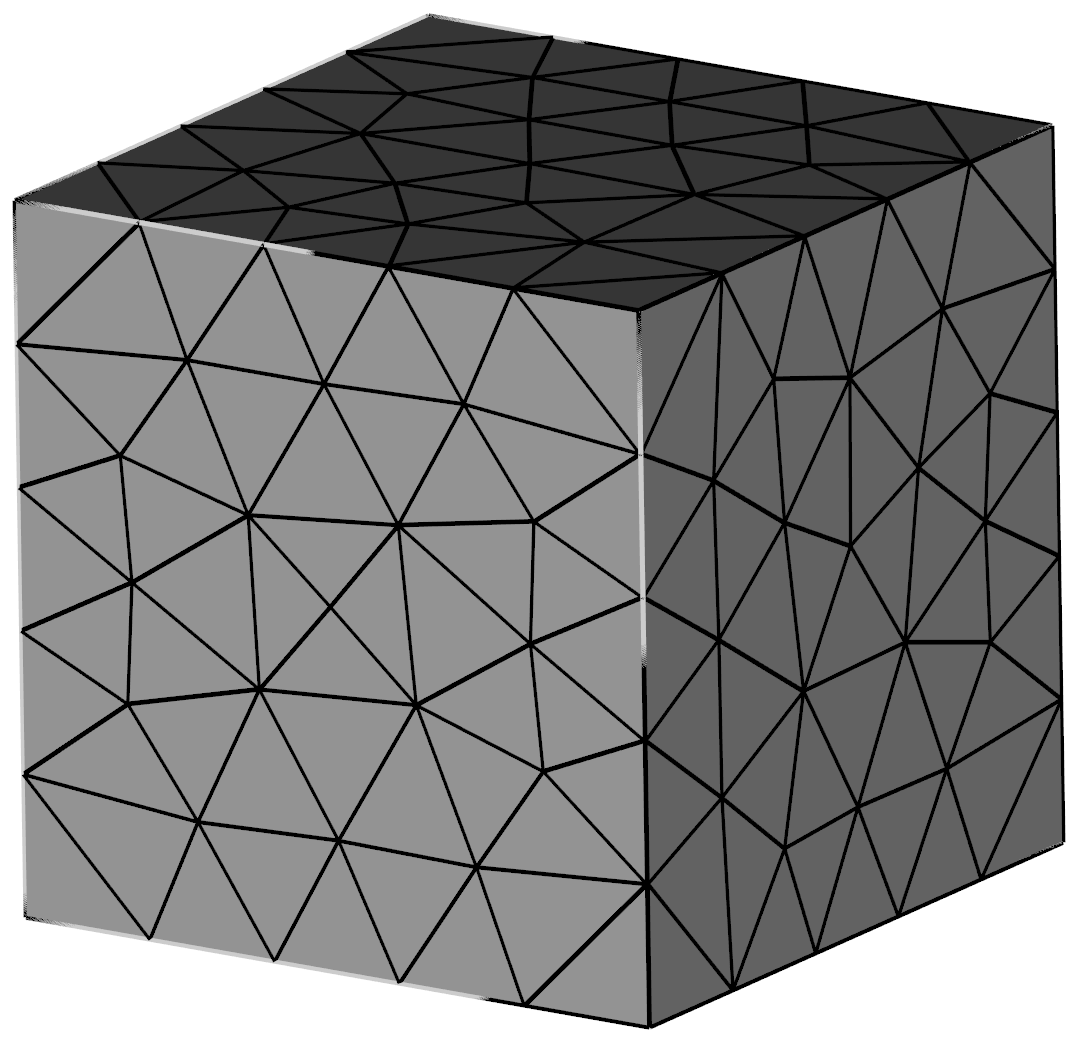}
    \caption{Tetrahedra}
  \end{subfigure}
  \caption{\jvrev{Example mesh types}}
  \label{fig:meshes}
\end{figure}

We show results here for evaluating the curvature of a sphere, an ellipsoid, \jvrev{and a sinusoid}.
In each case below, we use a unit-sized physical domain $(-0.5,0.5)^3$, meshed with a regular Cartesian grid, distorted hexahedrons, and finally a tetrahedral mesh, \jvrev{examples of which are shown} in \figref{meshes}. The distorted hexahedral case is generated \jvrev{by} randomly displacing the regular hexahedral node positions by up to 10\% of the edge length, producing non-convex elements. In the case of the regular Cartesian grid, we show results alongside those from the traditional height function method \autocite{cummins_estimating_2005,francois_balanced-force_2006}. We define the relative curvature error \jvrev{for a mixed cell $i$ as}
\begin{equation}
  E_i = \frac{\kappa_i - \kappa^{\mathrm{ex}}_i}{\kappa^{\mathrm{ex}}_i}
\end{equation}
\jvrev{where $\kappa^{\mathrm{ex}}_i$ is the exact curvature for cell $i$. For surfaces without constant curvature, the exact curvature is defined using a point on the exact surface for each cell, and is discussed in more detail for our ellipsoid test case in \secref{ellipsoid}.}
\jvrev{The $L_\infty$ norm is then the maximum absolute value of the error across all mixed cells in our domain. The $L_2$ norm is defined similarly to that used in previous literature \autocite{ivey_accurate_2015,ito_high-precision_2014,owkes_mesh-decoupled_2015,evrard_estimation_2017},}
\begin{equation}
  L_2 = \sqrt{\frac{\sum_i E_i^2 V_i}{\sum_i V_i}}
\end{equation}
where $V_i$ is the volume of cell $i$ \jvrev{and the sums are taken only over mixed cells.} A cell is taken to be mixed if its volume fraction is between $\delta$ and $1-\delta$, here taken to be $\delta = 10^{-5}$. We then plot the error norms against $N^{-1/3}$, where $N$ is the total number of cells in our mesh, loosely representing cell \jvrev{width for unstructured meshes}. For our regular mesh, this is equal to $\Delta x$.

\jvrev{For converging curvature calculations, an accurate volume fraction field is essential.} Volume fractions are initialized using a recursive adaptive refinement technique, similar to that used by \autocite{ivey_accurate_2015,cummins_estimating_2005,lopez_improved_2009} and briefly described here. Our material shape is described by a signed distance function to the interface. We say that a cell is mixed if not all of its node positions are on the same side of the interface. Pure cells have a volume fraction of 0 or 1, depending on whether they are inside or outside the shape. A mixed cell is subdivided
\jvrev{into tetrahedra by introducing nodes at the cell-center, face-centers, and edge-centers.}
The parent volume fraction is set to the volume-weighted average of its child volume fractions. This procedure is recursively applied for the children, down to a depth chosen here as 5 levels. At the finest level, mixed \jvrev{subcell} volume fractions are calculated from a piecewise-linear interface reconstruction generated from the signed distance function.

\subsection{Sphere}
\begin{figure}[tp]
  \centering
  \begin{subfigure}{0.3\textwidth}
    \includegraphics[width=0.98\textwidth]{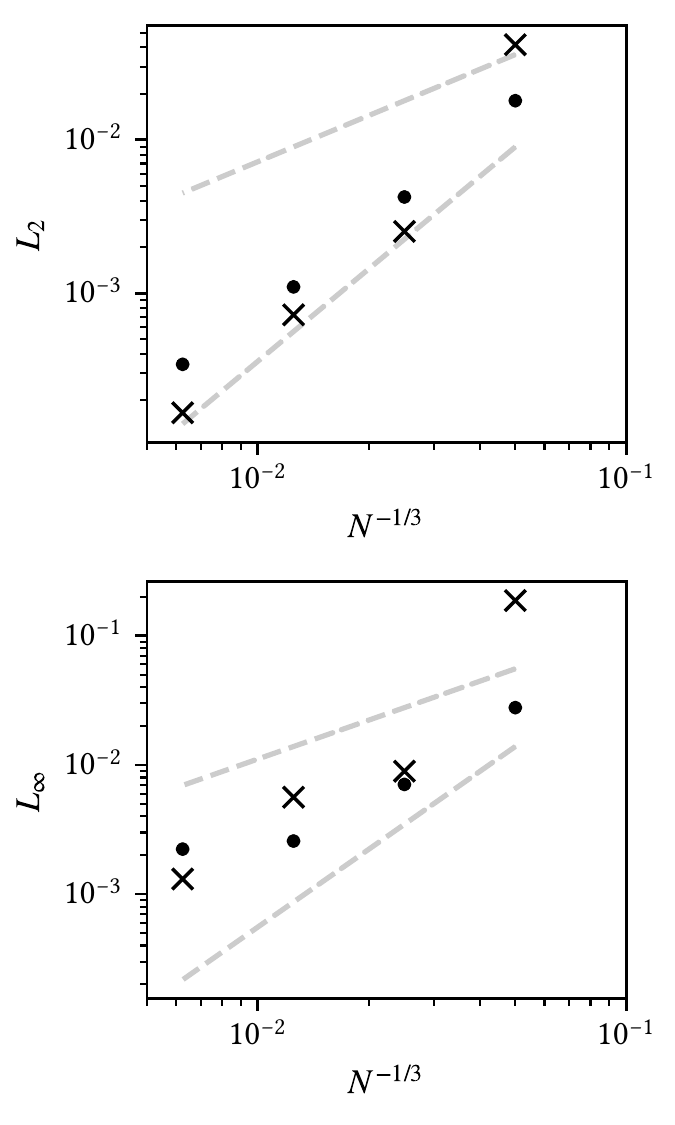}
    \caption{$L_2$ and $L_\infty$ norms on regular \jvrev{hexahedral} meshes}
  \end{subfigure}
  \hspace{1em}
  \begin{subfigure}{0.3\textwidth}
    \includegraphics[width=0.98\textwidth]{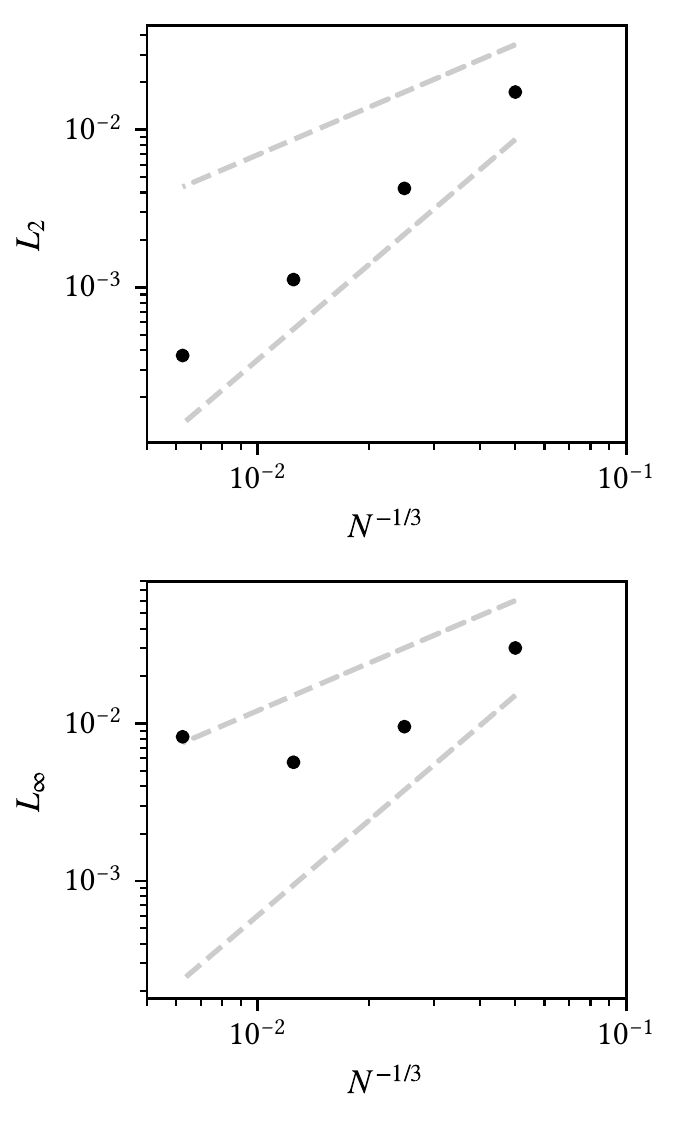}
    \caption{$L_2$ and $L_\infty$ norms on distorted hexahedral meshes}
  \end{subfigure}
  \hspace{1em}
  \begin{subfigure}{0.3\textwidth}
    \includegraphics[width=0.98\textwidth]{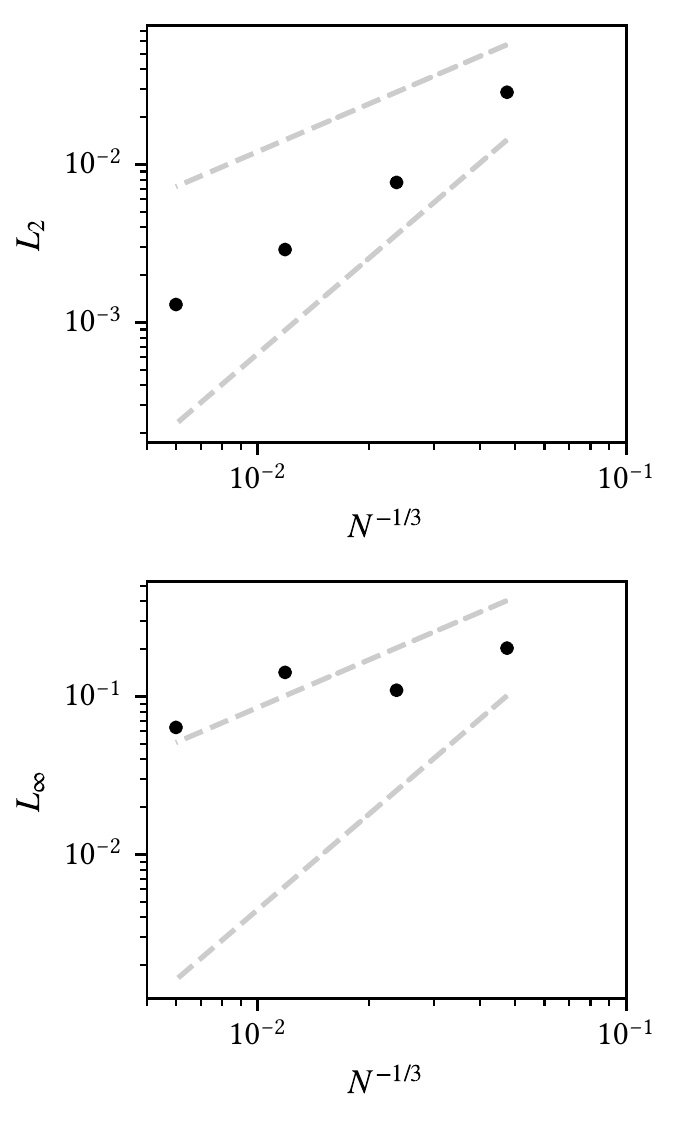}
    \caption{$L_2$ and $L_\infty$ norms on tetrahedral meshes}
  \end{subfigure}
  \caption{Grid refinement study for curvature of a sphere. Height function results are given by crosses, when applicable, our fitting approach results are given by dots, and first and second order convergence rates are shown with dashed lines.}
  \label{fig:sphere_conv}
\end{figure}

We initialize our domain with a sphere of radius $R = 0.35$ and centered at $x = y = 0$, such that the exact curvature is $\kappa_{\mathrm{ex}} = - 2 \,/\, R$. The results are shown in \figref{sphere_conv} \jvrev{and \tabref{sphere_conv}}. We find that on regular meshes, the fitting method produces errors similar to that of the height function method. It converges at second order, with the $L_\infty$ norm leveling off at higher resolutions. \jvrev{For the distorted hex meshes, we also find a second order convergence rate in the $L_2$ norm. For the $L_\infty$ norm we find initially second order convergence rates which then level off and finally reverse at the finest mesh. The case of tetrahedral meshes shows a second order convergence rate in the $L_2$ norm with very slight tapering at the finest mesh levels, and seemingly first order $L_\infty$ norm behavior. The most closely comparable method, the embedded height function method, converges in the $L_1$ norm at second order on regular meshes and better than first order on tetrahedral meshes for a similar problem \autocite{ivey_accurate_2015}.}

\begin{figure}[tp]
  \centering
  \includegraphics[width=0.8\textwidth]{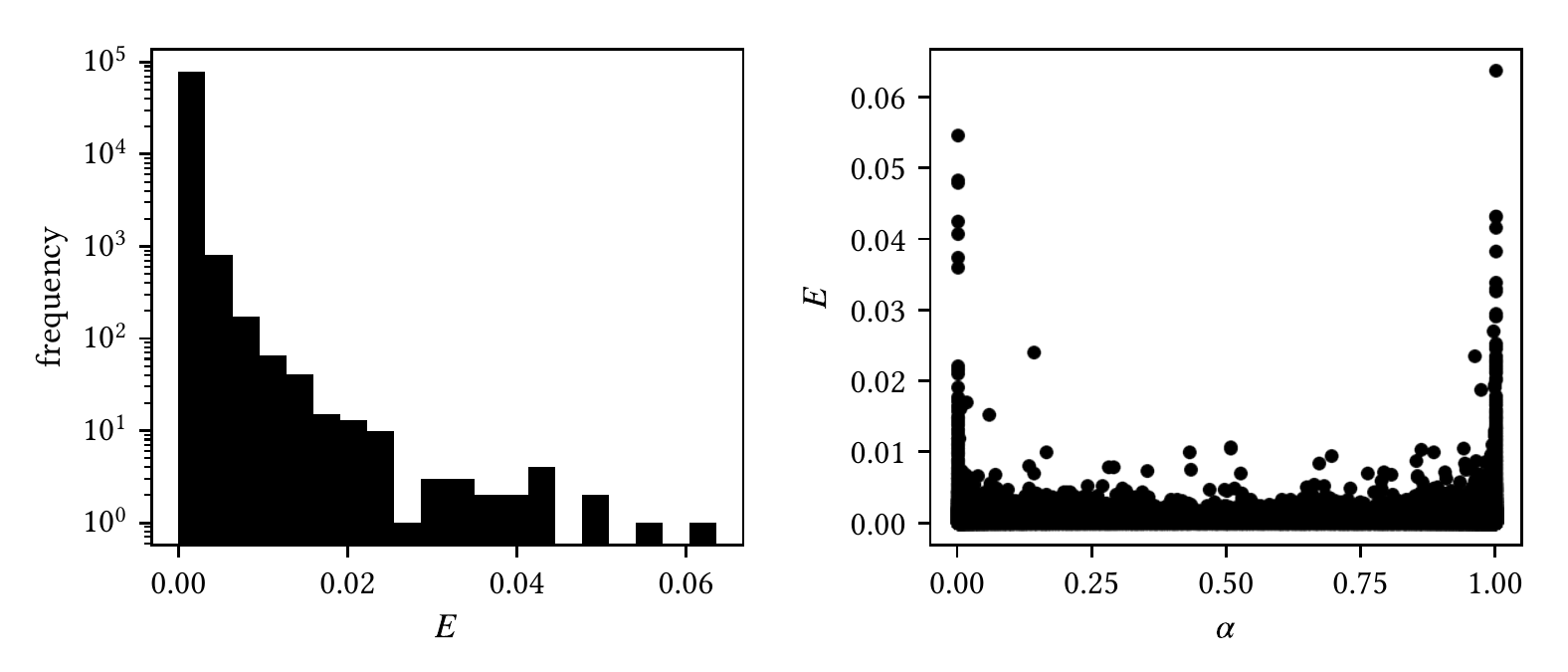}
  \caption{Histogram of the error \jvrev{(left), and error vs volume fraction for every cell (right)}. Shown for the case of a sphere on the finest tetrahedral mesh.} \label{fig:errhist}
\end{figure}

\jvrev{The $L_\infty$ norm is dominated by a few cells, shown in \figref{errhist}. \jvrev{Considering} the case of the finest tetrahedral mesh, for example, only 20 out of 80025 cells have an error greater than 0.025. We also see that these errors are clustered in cells with volume fractions very close to 0 or 1, where the interface reconstruction is very sensitive to small errors in the normal vector. A similar plot can be made for error against interface polygon area, which shows high error cells to cluster in cells with low interface polygon area.}

\begin{table}[tp]
  \centering
  \caption{Sphere Test Case Results}\vspace{-5pt} \label{tab:sphere_conv}
  \begin{tabular}{ ccc | ccc | ccc }
    \toprule
    \multicolumn{3}{c|}{Regular Hex Meshes} & \multicolumn{3}{c|}{Distorted Hex Meshes} & \multicolumn{3}{c}{Tetrahedral Meshes} \\
    \midrule
    $N^{-1/3}$ & $L_2$ & Order  &  $N^{-1/3}$ & $L_2$ & Order  &   $N^{-1/3}$ & $L_2$ & Order \\
    \hline
    5.00e-2 & 1.80e-2 &      & 5.00e-2 & 1.74e-2 &      & 4.75e-2 & 2.86e-2 &     \\
    2.50e-2 & 4.23e-3 & 2.1  & 2.50e-2 & 4.25e-3 & 2.0  & 2.38e-2 & 7.67e-3 & 1.9 \\
    1.25e-2 & 1.10e-3 & 1.9  & 1.25e-2 & 1.12e-3 & 1.9  & 1.19e-2 & 2.89e-3 & 1.4 \\
    6.25e-3 & 3.43e-4 & 1.7  & 6.25e-3 & 3.69e-4 & 0.8  & 5.99e-3 & 1.30e-3 & 1.2 \\
    \midrule
    $N^{-1/3}$ & $L_\infty$ & Order  &  $N^{-1/3}$ & $L_\infty$ & Order  &   $N^{-1/3}$ & $L_\infty$ & Order \\
    \hline
    5.00e-2 & 2.78e-2 &      & 5.00e-2 & 3.02e-2 &      & 4.75e-2 & 2.02e-1 &     \\
    2.50e-2 & 7.07e-3 & 2.0  & 2.50e-2 & 9.55e-3 & 1.7  & 2.38e-2 & 1.09e-1 & 0.9 \\
    1.25e-2 & 2.58e-3 & 1.5  & 1.25e-2 & 5.67e-3 & 0.8  & 1.19e-2 & 1.42e-1 & -0.4\\
    6.25e-3 & 2.23e-3 & 0.2  & 6.25e-3 & 8.23e-3 & -0.5 & 5.99e-3 & 6.36e-2 & 1.2 \\
    \bottomrule
  \end{tabular}
\end{table}

\subsection{Ellipsoid} \label{sec:ellipsoid}
We initialize our domain with an ellipsoid, defined by
\begin{equation}
  \label{eq:ellipsoid}
  \frac{x^2}{a^2} + \frac{y^2}{b^2} + \frac{z^2}{c^2} \le 1
\end{equation}
We choose $a = 0.35$, $b = 0.3$, and $c = 0.2$. The exact curvature for a cell is then evaluated using a height-function-like approach. We orient in the \jvrev{direction $\nv{x}$, $\nv{y}$, or $\nv{z}$} closest to the calculated normal, and evaluate the analytic expression for curvature from the explicit form of the interface, e.g.,
\begin{align}
  z &= c \sqrt{1 - \frac{x^2}{a^2} - \frac{y^2}{b^2}} \\
  \kappa^{\mathrm{ex}} &= -\frac{z_{xx} + z_{yy} + z_{xx}z_{y}^2 + z_{yy}z_{x}^2 - 2z_{xy}z_{x}z_{y}}{\left( 1 + z_x^2 + z_y^2\right)^{3/2}}
                    \label{eq:curvaturez}
\end{align}
where the curvature is evaluated at the $x$ and $y$ corresponding to cell center.

\begin{figure}[tp]
  \centering
  \begin{subfigure}{0.3\textwidth}
    \includegraphics[width=0.98\textwidth]{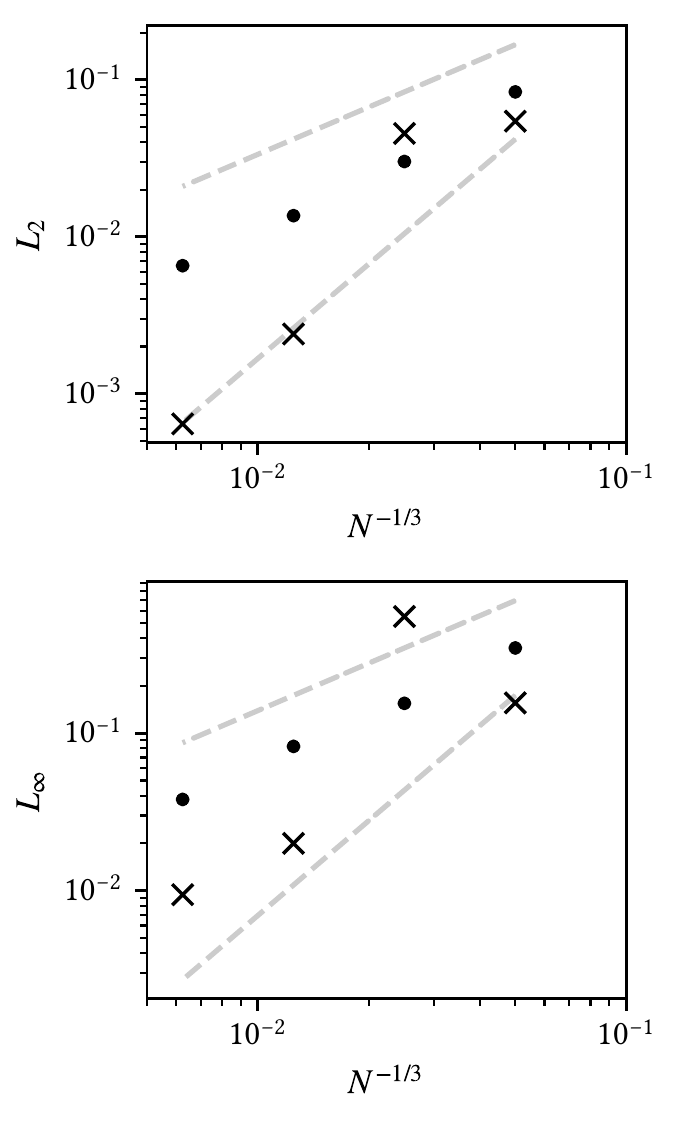}
    \caption{$L_2$ and $L_\infty$ norms on regular \jvrev{hexahedral} meshes}
  \end{subfigure}
  \hspace{1em}
  \begin{subfigure}{0.3\textwidth}
    \includegraphics[width=0.98\textwidth]{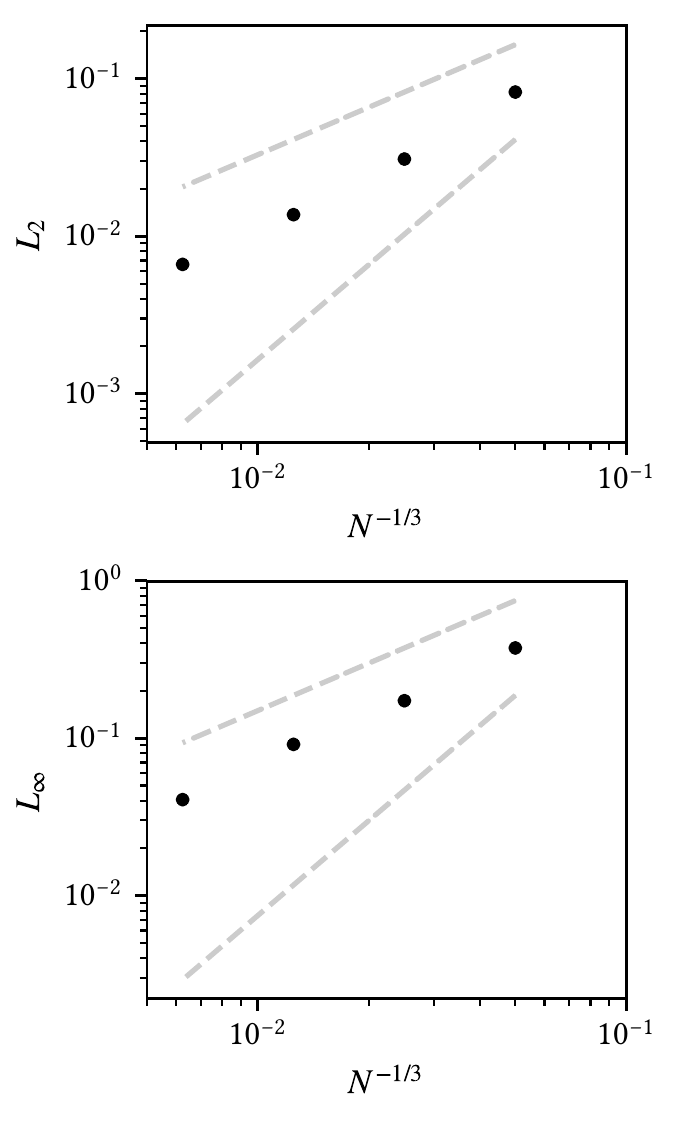}
    \caption{$L_2$ and $L_\infty$ norms on distorted hexahedral meshes}
  \end{subfigure}
  \hspace{1em}
  \begin{subfigure}{0.3\textwidth}
    \includegraphics[width=0.98\textwidth]{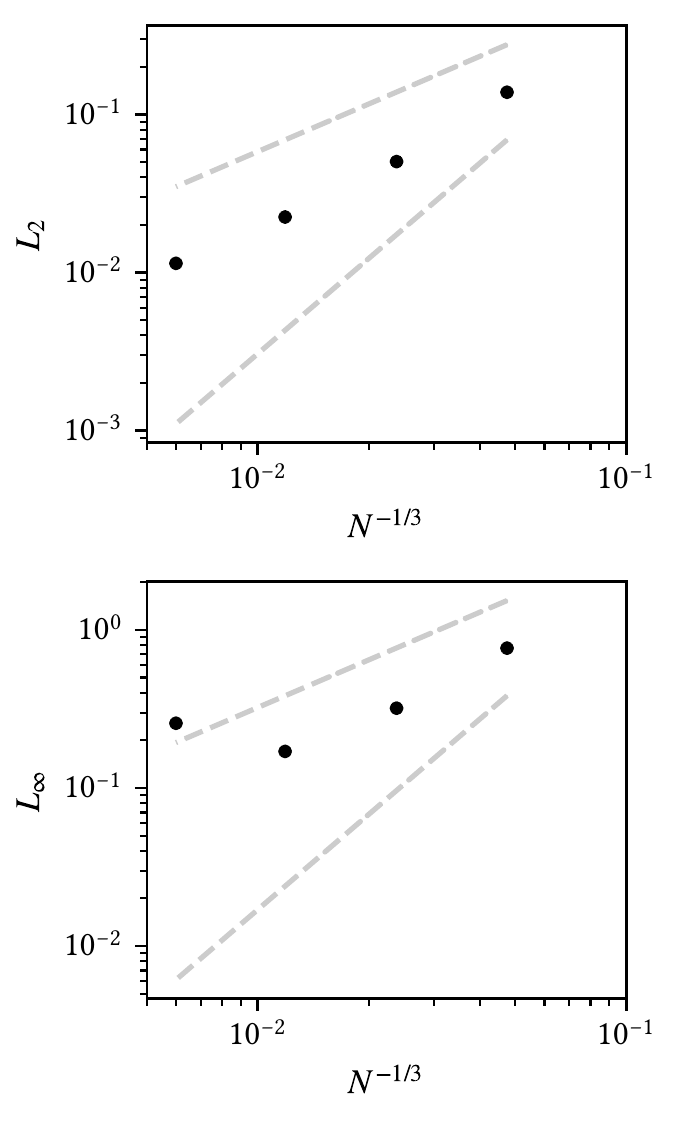}
    \caption{$L_2$ and $L_\infty$ norms on tetrahedral meshes}
  \end{subfigure}
  \caption{Grid refinement study for curvature of a ellipsoid. Height function results are given by crosses, when applicable, our fitting approach results are given by dots, and first and second order convergence rates are shown with dashed lines.}
  \label{fig:ellipsoid_conv}
\end{figure}

The results are shown in \figref{ellipsoid_conv} \jvrev{and \tabref{ellipsoid_conv}}. Here we find all norms to converge at first order, across all mesh types. \jvrev{On regular meshes, the errors are higher than that of the height function method, with the exception of an anomalous high error on one level.} The absolute errors are higher than for the sphere test, \jvrev{but the $L_\infty$ norm only increases in the finest tetrahedral mesh, and otherwise does not show signs of leveling off. Again in this case, \textcite{ivey_accurate_2015} report similar convergence rates in the $L_1$ norm for the embedded height function method for a similar problem. \textcite{evrard_estimation_2017} report second order convergence rates for a similar 2D problem, with simulated and exact curvatures evaluated using sub-cell quadrature of their fitted and exact surfaces.}

\begin{table}[tp]
  \centering
  \caption{Ellipsoid Test Case Results}\vspace{-5pt} \label{tab:ellipsoid_conv}
  \begin{tabular}{ ccc | ccc | ccc }
    \toprule
    \multicolumn{3}{c|}{Regular Hex Meshes} & \multicolumn{3}{c|}{Distorted Hex Meshes} & \multicolumn{3}{c}{Tetrahedral Meshes} \\
    \midrule
    $N^{-1/3}$ & $L_2$ & Order  &  $N^{-1/3}$ & $L_2$ & Order  &   $N^{-1/3}$ & $L_2$ & Order \\
    \hline
    5.00e-2 & 8.38e-2 &      & 5.00e-2 & 8.22e-2 &      & 4.75e-2 & 1.38e-1 &     \\
    2.50e-2 & 3.02e-2 & 1.5  & 2.50e-2 & 3.09e-2 & 1.4  & 2.38e-2 & 5.03e-2 & 1.5 \\
    1.25e-2 & 1.36e-2 & 1.1  & 1.25e-2 & 1.37e-2 & 1.2  & 1.19e-2 & 2.24e-2 & 1.2 \\
    6.25e-3 & 6.55e-3 & 1.1  & 6.25e-3 & 6.62e-3 & 1.0  & 5.99e-3 & 1.14e-2 & 1.0 \\
    \midrule
    $N^{-1/3}$ & $L_\infty$ & Order  &  $N^{-1/3}$ & $L_\infty$ & Order  &   $N^{-1/3}$ & $L_\infty$ & Order \\
    \hline
    5.00e-2 & 3.47e-1 &      & 5.00e-2 & 3.74e-1 &      & 4.75e-2 & 7.67e-1 &     \\
    2.50e-2 & 1.55e-1 & 1.2  & 2.50e-2 & 1.73e-1 & 1.1  & 2.38e-2 & 3.20e-1 & 1.3 \\
    1.25e-2 & 8.24e-2 & 0.9  & 1.25e-2 & 9.13e-2 & 0.9  & 1.19e-2 & 1.70e-1 & 0.9 \\
    6.25e-3 & 3.79e-2 & 1.1  & 6.25e-3 & 4.07e-2 & 1.2  & 5.99e-3 & 2.57e-1 & -0.6\\
    \bottomrule
  \end{tabular}
\end{table}

\subsection{\jvrev{Cosine Wave}}

\begin{figure}[tp]
  \centering
  \includegraphics[width=0.5\textwidth]{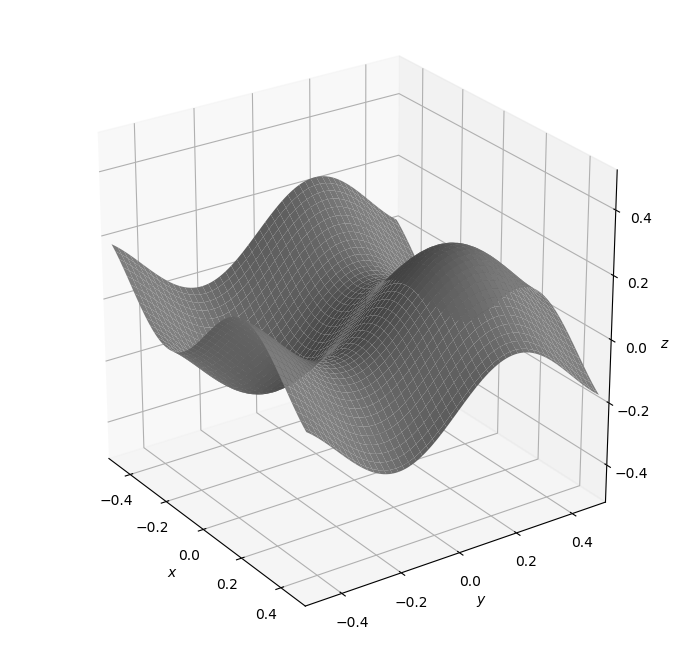}
  \caption{Cosine Wave}
  \label{fig:cosine_wave}
\end{figure}

\jvrev{Here we calculate the curvature for a cosine wave, given by}
\begin{equation} \label{eq:coswave}
  z \le A\left[\cos\!\left(\frac{2\pi}{L} \left(x-x_c\right)\right) + \cos\!\left(\frac{2\pi}{L} \left(y-y_c\right)\right)\right]
\end{equation}
\jvrev{We choose $A=1/8$, $L=4/5$, and $x_c = y_c = 1/5$. The surface is shown in \figref{cosine_wave}, and is noted for having regions of negative Gaussian curvature. The exact curvature is evaluated using \eqref{curvaturez}, using the $z=z(x,y)$ form of the interface from \eqref{coswave}. The maximum absolute exact curvature in this problem is $\abs{\kappa^{\mathrm{ex}}} \approx 15.42$. Since zero exact curvature is present in our domain, we use the absolute error rather than relative error for this problem.}
\begin{equation}
  E_i = \kappa_i - \kappa^{\mathrm{ex}}_i
\end{equation}

\begin{figure}[tp]
  \centering
  \includegraphics[width=0.98\textwidth]{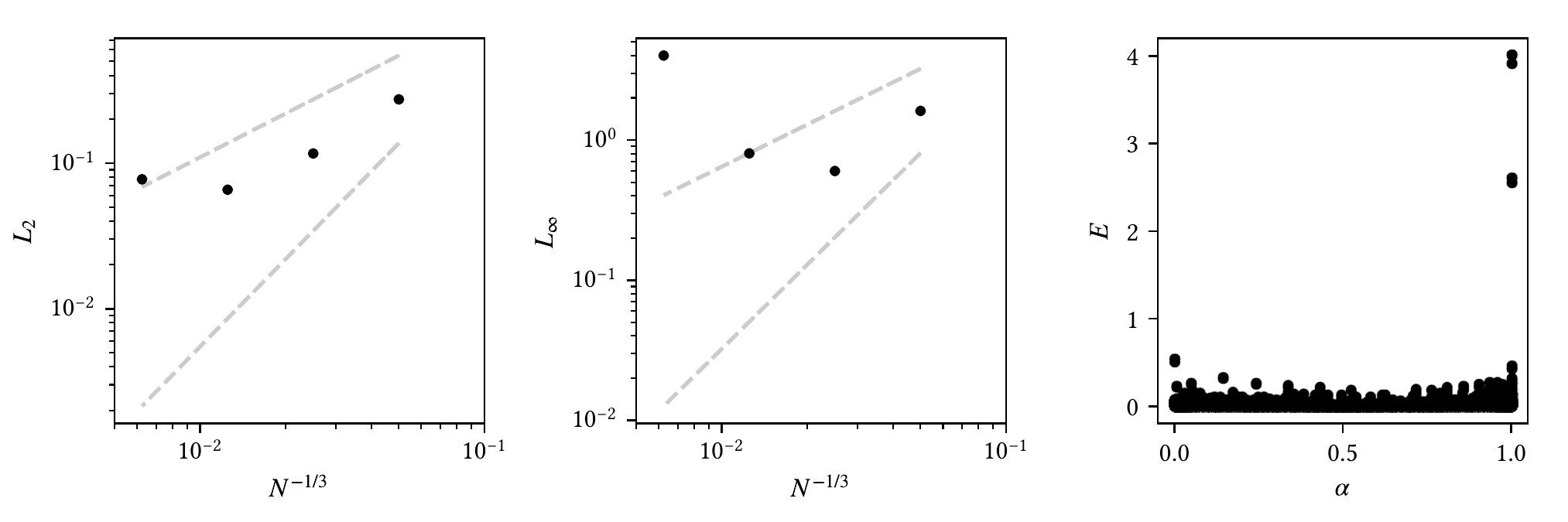}
  \caption{\jvrev{Results for the cosine wave test on regular hexahedral meshes, with $\delta=10^{-5}$. From left to right: $L_2$ norms, $L_\infty$ norms, and for the finest mesh the $E$-$\alpha$ graph for all mixed cells. Four high-error outlier cells are seen very close to $\alpha=1$.}}
  \label{fig:high_error}
\end{figure}

\begin{figure}[tp]
  \centering
  \begin{subfigure}{0.3\textwidth}
    \includegraphics[width=0.98\textwidth]{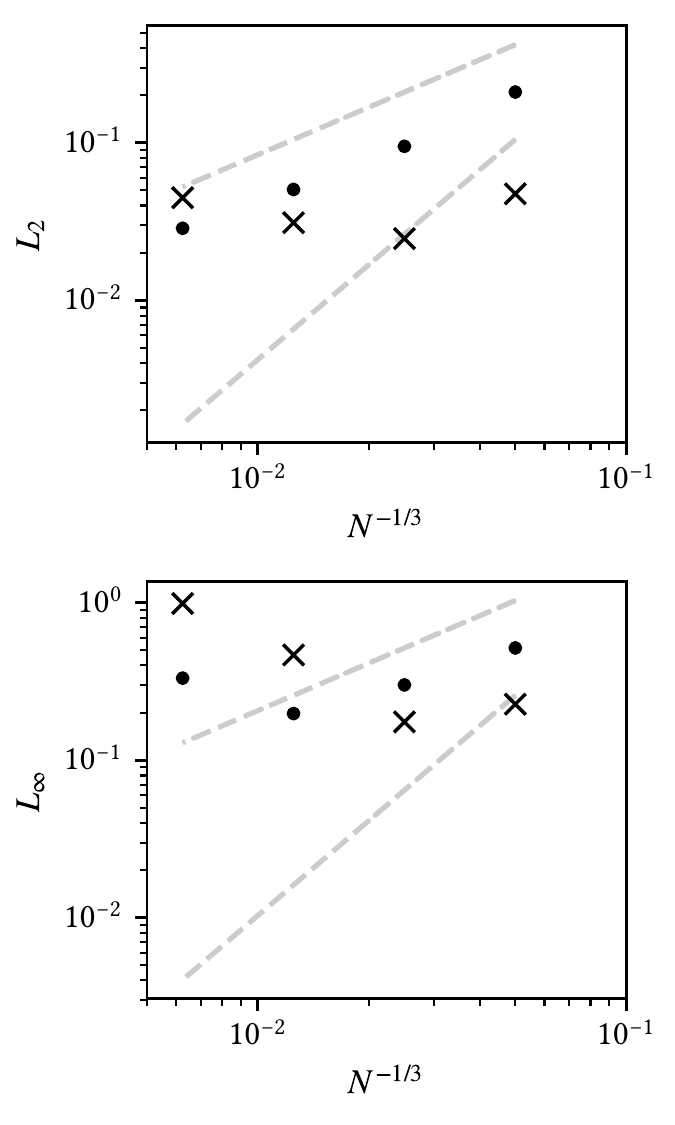}
    \caption{$L_2$ and $L_\infty$ norms on regular hexahedral meshes}
  \end{subfigure}
  \hspace{1em}
  \begin{subfigure}{0.3\textwidth}
    \includegraphics[width=0.98\textwidth]{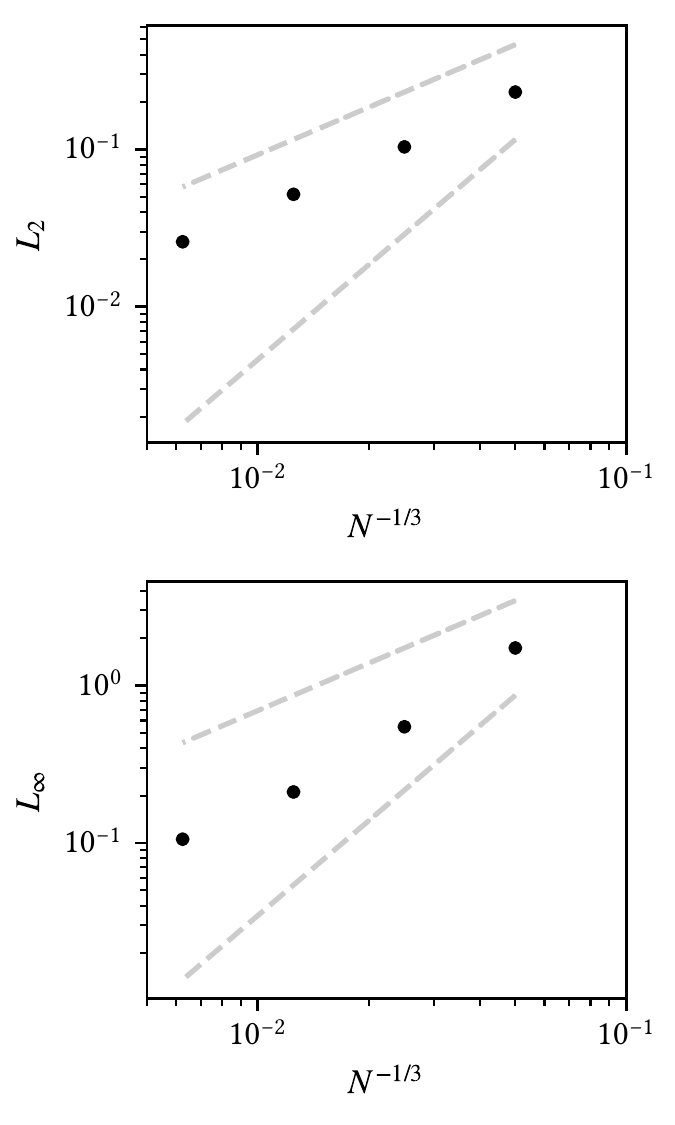}
    \caption{$L_2$ and $L_\infty$ norms on distorted hexahedral meshes}
  \end{subfigure}
  \hspace{1em}
  \begin{subfigure}{0.3\textwidth}
    \includegraphics[width=0.98\textwidth]{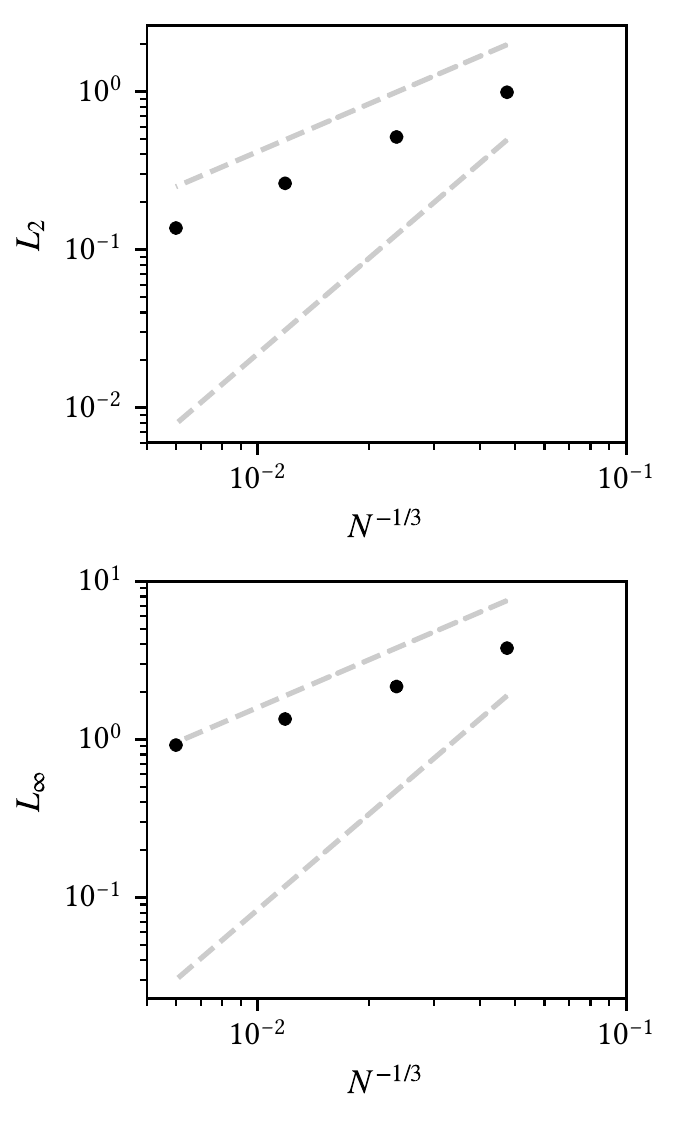}
    \caption{$L_2$ and $L_\infty$ norms on tetrahedral meshes}
  \end{subfigure}
  \caption{Grid refinement study for curvature of a cosine wave. Height function results are given by crosses, when applicable, our fitting approach results are given by dots, and first and second order convergence rates are shown with dashed lines.}
  \label{fig:cos_conv}
\end{figure}

\jvrev{For a regular mesh, we observe the results shown in \figref{high_error}. Here, the error is dominated by a few (<5) cells with volume fractions very close to zero or one, far more than in previous test cases. By changing our cutoff to $\delta = 10^{-3}$, the error norms better represent the overall curvature calculation quality. The results are shown in \figref{cos_conv} and \tabref{cos_conv}. We observe convergence behavior similar to the ellipsoid test case, with convergence rates near first order in each case, and beginning to level off at the finest resolutions. For the regular mesh, the $L_\infty$ norm increases at the finest mesh resolution. The traditional height function also exhibits nonconvergent behavior, and it produces higher errors than the paraboloid fitting approach at higher resolutions. On distorted hexahedral meshes, both $L_2$ and $L_\infty$ norms converge at first order. For the tetrahedral case, the $L_2$ norm converges at first order while the $L_\infty$ norm converges at less than first order.}

\begin{table}[tp]
  \centering
  \caption{Cosine Wave Test Case Results}\vspace{-5pt} \label{tab:cos_conv}
  \begin{tabular}{ ccc | ccc | ccc }
    \toprule
    \multicolumn{3}{c|}{Regular Hex Meshes} & \multicolumn{3}{c|}{Distorted Hex Meshes} & \multicolumn{3}{c}{Tetrahedral Meshes} \\
    \midrule
    $N^{-1/3}$ & $L_2$ & Order  &  $N^{-1/3}$ & $L_2$ & Order  &   $N^{-1/3}$ & $L_2$ & Order \\
    \hline
    5.00e-2 & 2.10e-1 &      & 5.00e-2 & 2.31e-1 &      & 4.75e-2 & 9.90e-1 &     \\
    2.50e-2 & 9.40e-2 & 1.1  & 2.50e-2 & 1.04e-1 & 1.2  & 2.38e-2 & 5.16e-1 & 0.9 \\
    1.25e-2 & 5.06e-2 & 0.9  & 1.25e-2 & 5.18e-2 & 1.0  & 1.19e-2 & 2.63e-1 & 1.0 \\
    6.25e-3 & 2.87e-2 & 0.8  & 6.25e-3 & 2.59e-2 & 1.0  & 5.99e-3 & 1.37e-1 & 1.0 \\
    \midrule
    $N^{-1/3}$ & $L_\infty$ & Order  &  $N^{-1/3}$ & $L_\infty$ & Order  &   $N^{-1/3}$ & $L_\infty$ & Order \\
    \hline
    5.00e-2 & 5.16e-1 &      & 5.00e-2 & 1.73    &      & 4.75e-2 & 3.77    &     \\
    2.50e-2 & 3.01e-1 & 0.8  & 2.50e-2 & 5.47e-1 & 1.7  & 2.38e-2 & 2.15    & 0.8 \\
    1.25e-2 & 1.98e-1 & 0.6  & 1.25e-2 & 2.11e-1 & 1.4  & 1.19e-2 & 1.34    & 0.7 \\
    6.25e-3 & 3.32e-1 & -0.7 & 6.25e-3 & 1.06e-1 & 1.0  & 5.99e-3 & 9.18e-1 & 0.6 \\
    \bottomrule
  \end{tabular}
\end{table}

\section{Summary}
We presented a novel method for interface curvature calculation, which involves reconstructing a \jvrev{paraboloid} surface from piecewise-linear interface reconstructions already calculated as part of the volume of fluid method. It is designed to work with a small stencil containing only cell neighbors sharing a node, making parallel implementations straightforward. Furthermore, it is applied to unstructured meshes, and is expected to handle mixed-element meshes with ease. We presented verification test cases on regular hex, distorted hex, and tetrahedral meshes for \jvrev{spherical, ellipsoid, and cosine wave} interface shapes. In all cases, we found the error to converge \jvrev{between first and} second order. \jvrev{Comparison to 2D results found by \textcite{evrard_estimation_2017} suggest that convergence rates may be improved by evaluating curvature using sub-cell quadrature rules rather than evaluating it nearest to the interface reconstruction centroid. This method naturally extends to larger stencils and higher order surfaces, which we expect would lead to higher order curvature calculations.} In future work, we will apply this method to curvature calculation necessary for modeling surface tension in an incompressible, balanced-force framework \autocite{francois_balanced-force_2006} and executed in parallel on distributed unstructured meshes.

All algorithms and tests for this paper can be found in the open-source code Pececillo, located at \url{https://gitlab.com/truchas/pececillo}.

\section*{Acknowledgments}
This work was carried out under the auspices of the National Nuclear Security Administration of the U.S. Department of Energy at Los Alamos National Laboratory under Contract No. DE-AC52-06NA25396, and supported by the Advanced Simulation and Computing Program.

Los Alamos Report LA-UR-17-30548.

\printbibliography[]

\end{document}